\documentclass[12pt]{article}

\usepackage{amsmath,amssymb,graphicx,multirow,xspace}
\usepackage[colorlinks=true,urlcolor=blue,anchorcolor=blue,citecolor=blue,filecolor=blue,linkcolor=blue,menucolor=blue,pagecolor=blue]{hyperref}
\usepackage[compress,numbers]{natbib}
\usepackage{placeins}
\usepackage{booktabs}
\usepackage{blindtext, rotating}
\usepackage{afterpage}
\usepackage{enumitem}
\usepackage{ marvosym }
\usepackage{verbatim}
\usepackage{authblk}

\usepackage{amsfonts}
\usepackage{float}
\usepackage{ccaption}
\usepackage{todonotes}
\usepackage{cancel}
\usepackage{verbatim}
\usepackage{color}
\usepackage{caption}
\usepackage{subcaption}
\usepackage{placeins}
\usepackage{fancyhdr}
\usepackage{epstopdf}
\usepackage{arydshln}
\usepackage{slashed}
\usepackage{psfrag}
\usepackage{epstopdf}
\usepackage{multicol}
\usepackage{enumerate}
\usepackage[top=1in,bottom=1in,left=1in,right=1in,footskip=0.5in,headheight=0.5in,footnotesep=0.2in,marginparwidth=0in,marginparsep=0in]{geometry}
\usepackage{titlesec}

\newcommand{\abs}[1]{\left| #1 \right|}

\newcommand{\bea}{\begin{eqnarray}\begin{aligned}}
\newcommand{\eea}{\end{aligned}\end{eqnarray}}

\newcommand{\CO}{\mathcal{O}}

\newcommand\fverb{\setbox\fverbbox=\hbox\bgroup\verb}
\newcommand\fverbdo{\egroup\medskip\noindent%
            \fbox{\unhbox\fverbbox}\ }
\newcommand\fverbit{\egroup\item[\fbox{\unhbox\fverbbox}]}
\newbox\fverbbox

\numberwithin{equation}{section}

\long\def\symbolfootnote[#1]#2{\begingroup%
\def\thefootnote{\fnsymbol{footnote}}\footnote[#1]{#2}\endgroup}

\newcommand{\be}{\begin{equation}}
\newcommand{\ee}{\end{equation}}

\newcommand{\mat}{\begin{pmatrix}}
\newcommand{\rix}{\end{pmatrix}}

\renewcommand{\bar}{\overline}

\newcommand{\beqa}{\begin{eqnarray}}
\newcommand{\eeqa}{\end{eqnarray}}

\newcommand{\beq}{\begin{equation}}
\newcommand{\eeq}{\end{equation}}

\numberwithin{equation}{section}

\begin{document}

\author[1]{Aria Basirnia}
\author[1]{Daniel Egana-Ugrinovic}
\author[2,3]{\authorcr Simon Knapen}
\author[1]{David Shih}

\affil[1]{ \small{New High Energy Theory Center, Rutgers University, Piscataway,
  NJ 08854}}
\affil[2]{Berkeley Center for Theoretical Physics,
University of California, Berkeley, CA 94720
}
\affil[3]{ Theoretical Physics Group, Lawrence Berkeley National Laboratory, Berkeley, CA 94720
}

\title{125 GeV Higgs from Tree-level $A$-terms}

\maketitle

\begin{abstract}

We present a new mechanism to generate large $A$-terms at tree-level in the MSSM through the use of  superpotential operators. The 
mechanism trivially resolves the $A/m^2$ problem which plagues models with conventional, loop-induced $A$-terms. We study both MFV and non-MFV models; in the former, naturalness motivates us to construct a UV completion using Seiberg duality. Finally, we study  the phenomenology  of these models when they are coupled to minimal gauge mediation. We find that after imposing the Higgs mass constraint, they are largely out of reach of LHC Run I, but  they will be probed at Run II. Their fine tuning is basically the minimum possible in the MSSM.

\end{abstract}

\newpage

\section{Introduction}

The discovery of the Higgs boson with a mass near 125~GeV \cite{Aad:2012tfa, Chatrchyan:2012ufa} has important consequences for physics beyond the Standard Model, especially supersymmetry. In the MSSM, it implies that the stops must either be very heavy or have a large trilinear coupling (``$A$-term") with the Higgs \cite{Hall:2011aa, Heinemeyer:2011aa, Arbey:2011ab, Arbey:2011aa, Draper:2011aa, Carena:2011aa, Cao:2012fz,  Christensen:2012ei, Brummer:2012ns}. The large $A$-term scenario is more interesting from several points of view. It is less fine-tuned and it allows for lighter ($\sim$~1~TeV) stops that are still within reach of the LHC. It also presents an interesting model-building challenge -- prior to the discovery of the Higgs, mechanisms for generating the $A$-terms from an underlying model of SUSY-breaking mediation were not well-explored.  

In the framework of gauge mediated SUSY-breaking (GMSB) (for a review and original references, see \cite{Giudice:1998bp}), the problem of how to obtain large $A$-terms becomes especially acute. In GMSB, the $A$-terms are always negligibly small at the messenger scale. If the messenger scale is sufficiently high and the gluino sufficiently heavy, a sizable weak scale $A$-term with relatively light stops may be generated through RG-running  \cite{Draper:2011aa}. However, this setup is in strong tension with electroweak symmetry breaking (EWSB)  \cite{GGMinprogress}. This strongly motivates extending gauge mediation with additional MSSM-messenger couplings
that generate $A$-terms through threshold corrections at the messenger scale. 

In all models for $A$-terms considered since the observation of a Higgs boson at 125 GeV \cite{Kang:2012ra,Craig:2012xp,Evans:2013kxa,Craig:2013wga,Knapen:2013zla,Abdullah:2012tq,Kim:2012vz,Byakti:2013ti,Calibbi:2013mka,Jelinski:2013kta,Galon:2013jba,Fischler:2013tva,Ding:2013pya,Calibbi:2014yha}, the focus has been on generating $A$-terms at one-loop level through weakly coupled messengers. Integrating out the messengers produces one or more of the following K\"ahler operators
\begin{equation}\label{AtermKahler}
 \frac{1}{16\pi^2 } {1\over M}  X^\dagger H_u^\dagger H_u\,\, , \quad\quad \frac{1}{16\pi^2 } {1\over M}  X^\dagger Q_3^\dagger Q_3\,\, ,\quad\quad  \frac{1}{16\pi^2 } {1\over M}   X^\dagger \overline{u}_3^\dagger \overline{u}_3
\end{equation}
Here $X$ is a field that spontaneously breaks SUSY, and $M$ is the messenger scale. After substituting $\langle X\rangle = \theta^2 F_X$ and integrating out the auxiliary components of the MSSM fields, one obtains the desired $A$-term 
\begin{equation}
\mathcal{L}\supset y_t A_t H_u Q_3\bar{u}_3 \quad , \quad  A_t \sim {1 \over 16\pi^2}{F_X\over M} 
\end{equation} 
This setup has the advantage that the $A$-terms come out parametrically the same size as the other soft masses in GMSB (one-loop gaugino masses, two-loop scalar mass-squareds). However, one-loop $A$-terms from (\ref{AtermKahler}) introduce a host of complications as well. First and foremost is the ``$A/m^2$ problem" \cite{Craig:2012xp}: in addition to the $A$-terms, one also generates a scalar mass-squared at one-loop, completely analogous with the more well-known $\mu/B_\mu$ problem. A one-loop scalar mass-squared would overwhelm the GMSB contributions and lead to serious problems with fine-tuning and/or EWSB. Previous solutions to the $A/m^2$ problem include taking the messengers to be those of minimal gauge mediation \cite{Craig:2012xp}, or having the hidden sector be a strongly-coupled SCFT \cite{Craig:2013wga,Knapen:2013zla}. 

In this paper, we will explore a new solution to the $A/m^2$ problem:  models where the $A$-terms are generated {\it at tree-level in the MSSM-messenger couplings}. The advantage with this approach is that there is simply no $A/m^2$ problem to begin with, since at worst any accompanying sfermion mass-squareds would be tree-level as well.  An added benefit of this approach is that it will lead us to a consider an interesting new operator for the $A$-terms: one which arises in the effective \emph{superpotential}, rather than in the K\"ahler potential. As we will see, this superpotential operator will have qualitatively different effects on the MSSM soft terms as compared to K\"ahler potential operators.

The basic setup is quite simple. To generate a tree-level $A$-term, either the Higgs or stops must mix with the messengers in the mass-matrix. For example, consider the superpotential
\begin{equation}\label{exampleH}
W =  X' H_u \tilde{\phi} + \lambda_u^{ij}\phi Q_i \overline{u}_j + M \tilde{\phi}\phi
\end{equation} 
Here $X'$ is another spurion for SUSY-breaking, and $\phi$, $\tilde\phi$ are heavy messenger fields. Upon integrating  out the messengers at the scale $M$, one generates the effective superpotential operator
\begin{equation}\label{eq:effsuperpotential}
W_{eff}\supset -{ \lambda_u^{ij}\over M  }X' H_u Q_i \overline{u}_j
\end{equation} 
Note that because of the SUSY non-renormalization theorem, $W_{eff}$ can {\it only} arise at tree-level, so it is perfectly suited for our purposes. In order to produce an $A$-term of the correct size, one must have\footnote{Note that this is a loop factor smaller than the usual GMSB relation. A smaller $F$-term satisfying this hierarchy can easily be dynamically generated using weakly-coupled messengers, see e.g.\ \cite{Komargodski:2008ax}. In this paper we will simply assume that $F_{X'}$ of the right size can be obtained somehow and not explore it any further.} 
\begin{equation}\label{eq:FXrel}
{F_{X'}\over M}\sim \CO({\rm TeV})
\end{equation}
 The tree-level $A$-term originating from (\ref{eq:effsuperpotential}) is minimally flavor violating (MFV), provided that the operator in (\ref{eq:effsuperpotential})  generates the full up-type Yukawa coupling of the MSSM. For this to work, $X'$ should acquire a lowest component vev of size $\sim M$.

The interesting complication in these models comes from the fact that when integrating out the messengers, in addition to the superpotential operator  (\ref{eq:effsuperpotential}), a K\"ahler potential operator is also generated at tree-level. For example, in the model (\ref{exampleH}), one generates the term:
\begin{equation} \label{listofmass}
K_{eff} \supset {1\over M^2} X'^\dagger X' H_u^\dagger H_u  
\end{equation} 
(For a more general treatment of the K\"ahler operators, see appendix A.)
This leads to a soft mass for $H_u$ of roughly the same order as the $A$-term:
\begin{equation}\label{littleamhexample}
\delta m_{H_u}^2 = - \frac{y_t^2}{|\lambda_u^{33}|^2} |A_t|^2 
\end{equation}
For $\lambda_u^{33}\lesssim 1$, this represents a large, irreducible contribution to $m_{H_u}^2$, and correspondingly to the fine-tuning of the electroweak scale. 
This is another manifestation of the ``little $A/m^2$ problem" encountered in \cite{Craig:2012xp}, whereby a large $A$-term was accompanied by an equally large sfermion mass-squared. In \cite{Craig:2012xp}, the situation was even worse, because the contribution was irreducible with a fixed coefficient:
\begin{equation}\label{littleamh}
\delta m_{H_u}^2 = \abs{A_t}^2 
\end{equation} 
There both the $A$-terms and the irreducible contribution to $m_{H_u}^2$ (\ref{littleamh}) originated from integrating out the auxiliary components of the MSSM fields in the   first K\"ahler operator in (\ref{AtermKahler}). 
Since we are starting instead with the effective superpotential operator (\ref{eq:effsuperpotential}), the coefficient in (\ref{littleamhexample}) is free to vary in our present models. Importantly, however, we will see that the sign in (\ref{littleamhexample}) is always \emph{negative}, such that (\ref{littleamhexample}) does not jeopardize electroweak symmetry breaking, in contrast to the relation in (\ref{littleamh}).

In this paper, we will consider various ways to alleviate the fine-tuning problem introduced by the little $A/m^2$ problem (\ref{littleamhexample}). Clearly, if $\lambda_{33}$ is taken to be large (e.g.\ $\lambda_{33}\sim 3$), then the little $A/m^2$ problem is ameliorated. This requires a UV completion at a relatively low scale. We will provide such a UV completion in this paper, using a novel application of Seiberg duality \cite{Seiberg:1994pq, Seiberg:1994bz}. 

Alternatively, one can consider non-MFV models obtained from (\ref{exampleH}) by exchanging the role played by $H_u$ with $\bar u_3$:\footnote{Because these models are not MFV, one should worry about the potential constraints from precision flavor and CP observables. This is beyond the scope of this work (see however \cite{flavorinprogress}). We will assume for simplicity (as in  \cite{Evans:2013kxa}) that the coupling $\kappa$ is real and fully aligned with the third generation. We will also focus on the $\bar u_3$ model because then the flavor violation is limited to the up-squark sector and the constraints are much weaker.}
\begin{equation}
W = X' \bar{u}_3  \tilde{\phi}_u + \kappa H_u Q_3 \phi_u + M \tilde{\phi}_u \phi_u\label{exampleU}
\end{equation}
 For this model the expression analogous to (\ref{littleamhexample}) contains $m_{\bar u_3}^2$ instead of $m_{H_u}^2$. As in \cite{Evans:2013kxa}, the fine-tuning is greatly reduced with respect to the perturbative MFV case because the stop contribution to $m_{H_u}^2$ is diluted by a loop factor. Moreover, the situation is even better than in \cite{Evans:2013kxa}, because in that case there were still sizeable two-loop contributions to $m_{H_u}^2$, whereas here the contribution is solely to the squarks.

An important thing to note about the framework for generating tree-level  $A$-terms presented in this paper is that it can in principle be tacked on to any mediation mechanism for the rest of the MSSM soft terms; the framework itself does not lead to a particularly compelling choice. This is in contrast to the one-loop models considered previously, whereby the $A$-term messengers also contributed to the MSSM soft spectrum through minimal gauge mediation, and thus GMSB was the most economical choice. Moreover, the tree-level $A$-term module does not affect the overall phenomenology much; the  one essential difference occurs in the non-MFV models, where the stops can be split by several TeV due to the non-MFV analogue of (\ref{littleamh}). 

For simplicity and concreteness, in this paper we will couple our models to minimal gauge mediation (MGM) \cite{Dine:1993yw,Dine:1994vc,Dine:1995ag}. We will see that after imposing the Higgs mass constraint, the models are typically out of reach of Run I LHC; however they will be accessible (especially the lightest stop) at 14 TeV LHC. Finally, we will estimate the fine tuning in these models and show that they achieve essentially the best tuning possible in the MSSM (percent level). 

The remainder of this paper is organized as follows: Since no strongly coupled UV completion is needed for the non-MFV models, we discuss those first in section 2, as well as their phenomenology when coupled to minimal gauge mediation. In section 3 we analyze the MFV example in a similar way. In section 4, we UV complete  the MFV model using Seiberg duality. Finally, in the conclusions we list some potential future directions suggested by our work. A general discussion of the little $A/m^2$ problem and Landau poles in models for tree-level $A$-terms  is left for appendix \ref{app:generalization}. 

\section{A non-MFV model}
\label{sec:NMFV models}

As discussed in the introduction, the non-MFV model (\ref{exampleU}) has a less severe version of the little $A/m^2$ problem, and thus does not need an immediate UV completion, unlike the MFV model (\ref{exampleH}). Since the story is simpler here,  let us start by analyzing the non-MFV model in detail. Apart from the issues of flavor alignment discussed in the introduction, the form of the renormalizable superpotential (\ref{exampleU}) is the most general that couples the spurion, messengers and MSSM fields up to terms that are irrelevant for our purposes (powers of the spurion $X'$ and a small soft mass for the messenger pair from $X'\phi_{u}\tilde{\phi}_{u}$). 

After diagonalizing the mass matrix and integrating out $\phi_u, \tilde{\phi}_u$ at the messenger scale $M$, we obtain the IR effective theory
\bea
W_{eff}&\supset -\kappa \frac{X'}{M} H_u Q_3 \bar{u}_3\\
K_{eff}&\supset \frac{X^{'\dagger} X'}{M^2} \bar{u}_3^\dagger \bar{u}_3
+\frac{\kappa^2}{M^2} H_u^\dagger H_u Q_3^\dagger  Q_3
\label{eq:NMFVeff}
\eea
The irrelevant operator induced in the low energy superpotential leads to an $A$-term for the corresponding MSSM fields after substituting $\langle X'\rangle = \theta^2 F_{X'}$. However, an additional contribution to $m_{\bar{u}_3}^2$ from the first term in the K\"ahler potential is also induced, such that
\begin{equation}
\delta m_{\bar{u}_3}^2 = -\frac{y_t^2}{\kappa^2} A_t^2 
\label{eq:littleAmu}
\end{equation}
Note that the contribution to $m_{\bar{u}_3}^2$ is negative, so to avoid a tachyonic right handed stop, it must be cancelled off by additional contributions at the messenger scale (e.g.\ from GMSB) or from MSSM renormalization group running from the messenger scale down to the weak scale. If $\kappa\sim1$, the fine tuning from (\ref{eq:littleAmu}) is comparable to the fine tuning from the $A$-term itself, since both enter the running of $m_{H_u}^2$ in exactly the same fashion. Taking $\kappa >1$ therefore does not substantially improve the overall fine tuning of the model. One major improvement relative to the non-MFV models considered in  \cite{Evans:2013kxa} is that there are no sizeable contributions generated to $m_{H_u}^2$ from integrating out the messengers.

To study the phenomenology of a model with tree-level $A$-terms and a 125~GeV Higgs, we must add our tree-level $A$-term module (\ref{exampleU}) to an underlying model for the rest of the MSSM soft masses. While in principle any model could be used, GMSB is a particularly well-motivated choice given the SUSY flavor problem. So for simplicity and concreteness, let us now specialize to the case of minimal gauge mediation (MGM) with ${\bf 5}\oplus {\bf \bar 5}$ messengers \cite{Dine:1993yw,Dine:1994vc,Dine:1995ag}. 

The parameter space of our model is as follows. The MGM sector of the model is characterized by four parameters: messenger index $N_{m}$,  $\tan{\beta}$, messenger scale $M$ and SUSY-breaking mass scale $\frac{F_X}{M}$, where $F_X$ is the highest component vev of the SUSY breaking spurion. We take the masses of the additional messengers in (\ref{exampleU}) to be the same scale $M$ for simplicity. We consider $\mu$ and $B_{\mu}$ to be determined by the EWSB conditions and we remain agnostic about their origin. Finally, our model contains additional parameters $\frac{F_{X'}}{M}$, which sets the scale for the tree level contribution to $A_t$, and the coupling $\kappa$ (see  (\ref{exampleU})).

 A low messenger scale $M=250$ TeV and a large messenger number $N_m=3$ are motivated by the simultaneous requirements of reducing the tuning from the RG while allowing a large enough SUSY scale to be achieved for the Higgs mass. (A different choice of messenger number does  not alter the phenomenology heavily, for reasons that will be explained later.) We take $\tan\beta = 20$ to saturate the tree level bound of the Higgs mass and $\kappa=1$  for simplicity and perturbativity.  With these choices, the parameter space of our models reduces to $(\frac{F_{X'}}{M}, \frac{F_{X}}{M})$. (Recall that we must take $\frac{F_{X'}}{M}\sim\frac{1}{16\pi^2}\frac{F_{X}}{M}$ to achieve $A$-terms comparable to the GMSB soft masses.) To make contact with the IR observables, we can trade $(\frac{F_{X'}}{M}, \frac{F_{X}}{M})$ by the IR values of $A_t$ and the mass of the lightest stop $m_{\tilde{t}_1}$ or the mass of the lightest stau $m_{\tilde{\tau}_1}$. This parametrization is especially relevant for the LHC phenomenology, since $\tilde t_1$ and $\tilde \tau_1$ are the lightest colored particle and the NLSP respectively, as will be seen shortly. 

\begin{figure}
        \centering
        \begin{subfigure}[b]{0.45\textwidth}
                \includegraphics[width=\textwidth]{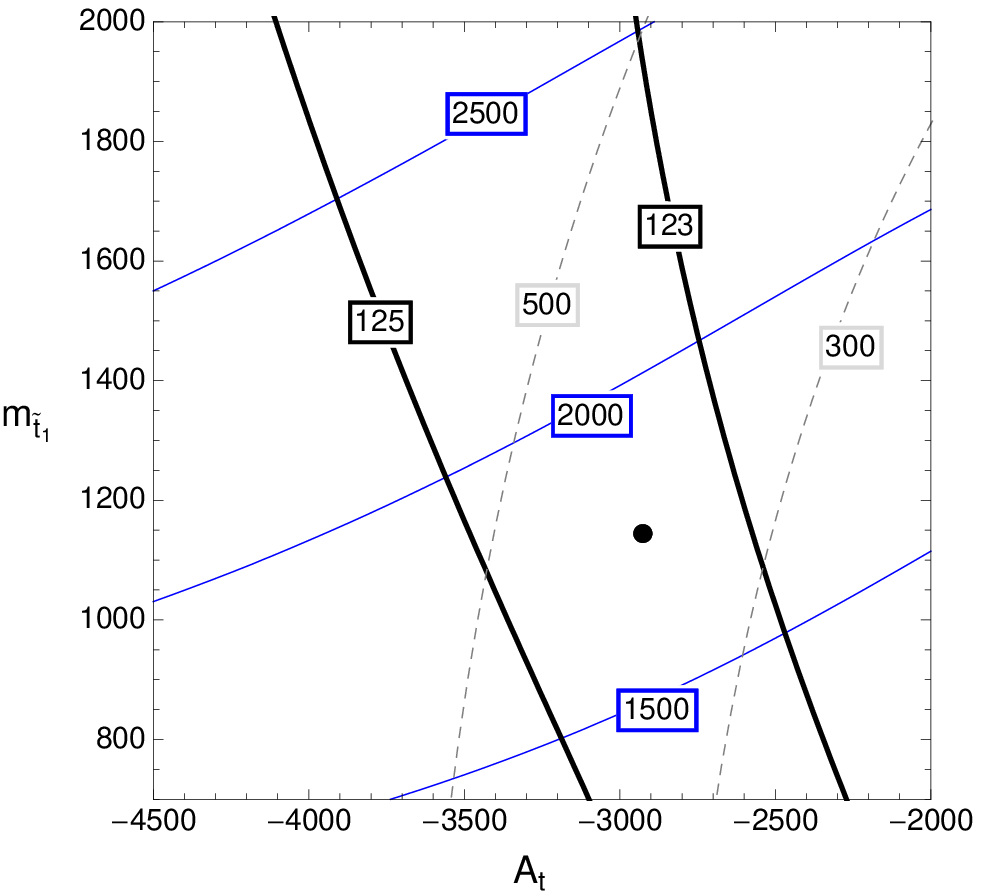}
                \label{fig:nmfvstop}
        \end{subfigure}%
        \hfill
        ~ 
        \begin{subfigure}[b]{0.45\textwidth}
                \includegraphics[width=\textwidth]{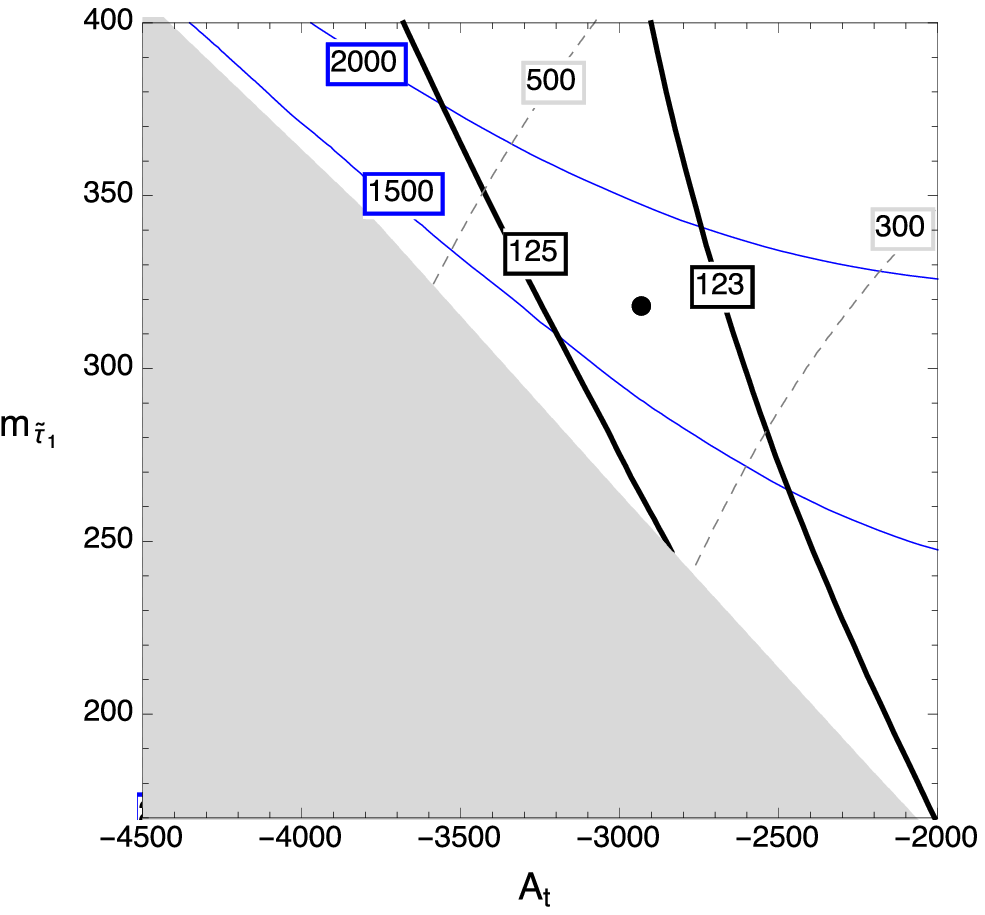}
                \label{fig:nmfvstau}
        \end{subfigure}
        \caption{Contours of the Higgs mass (black), geometric mean of the stop masses (blue) and tuning (dashed), in the  $(A_t,m_{\tilde{t}_1})$ (left) and $(A_t,m_{\tilde{\tau}_1})$ (right) planes. The shaded region on the  $(A_t, m_{\tilde{\tau}_1})$ plane corresponds to points with tachyonic stops. The black dot on both figures corresponds to the same point in parameter space, with a spectrum presented in figure \ref{fig:NMFVspec}. All quantities are evaluated at $M_{SUSY}$.}
 \label{fig:nmfv}
\end{figure}

To generate the IR spectrum we use \texttt{SOFTSUSY 3.5.1}  \cite{Allanach:2001kg}. Fine tuning $\Delta_{FT}$ is calculated according to the measure introduced in \cite{Evans:2013kxa}, given by
\bea
\label{eq:tuning}
&\Delta_i \equiv \frac{\partial \log m_z^2}{\partial \log \Lambda_i^2}&\Lambda_i \in \{g_1^2 \frac{F_{X}}{M}, g_2^2\frac{F_{X}}{M}, g_3^2\frac{F_{X}}{M},\frac{F_{X'}}{M}, \kappa \frac{F_{X'}}{M},\mu\}\\
&\Delta_{FT} \equiv \max \Delta_i. &
\eea
The results are presented in figure \ref{fig:nmfv} where we show contours of the Higgs mass, tuning and $M_{SUSY}$, both in the $( A_t,m_{\tilde{t}_1})$ and $(A_t, m_{\tilde{\tau}_1})$ planes. Note that $M_{SUSY}$ is significantly larger than $m_{\tilde{t}_1}$. This is because the two stop soft masses are split due to the negative contribution to $m_{\bar{u}_3}^2$ in (\ref{eq:littleAmu}). In the gray shaded region the GMSB contribution is insufficient to cancel this negative contribution, and the spectrum is invalidated by a  stop tachyon. The main source of tuning in this model is the running effect due to the colored spectrum or the $A$-term. From the Higgs and tuning contour lines in both figures, we see that the model is able to reproduce the Higgs mass, while keeping fine tuning to the percent level (which is basically the best that can be achieved in the MSSM). Moreover, the Higgs mass can be reproduced in interesting parts of parameter space, where there is both a light colored particle $m_{\tilde{t}_1}$ and a light slepton $m_{\tilde{\tau}_1}$.

\begin{figure}[t]
\begin{center}
\includegraphics[width=12cm]{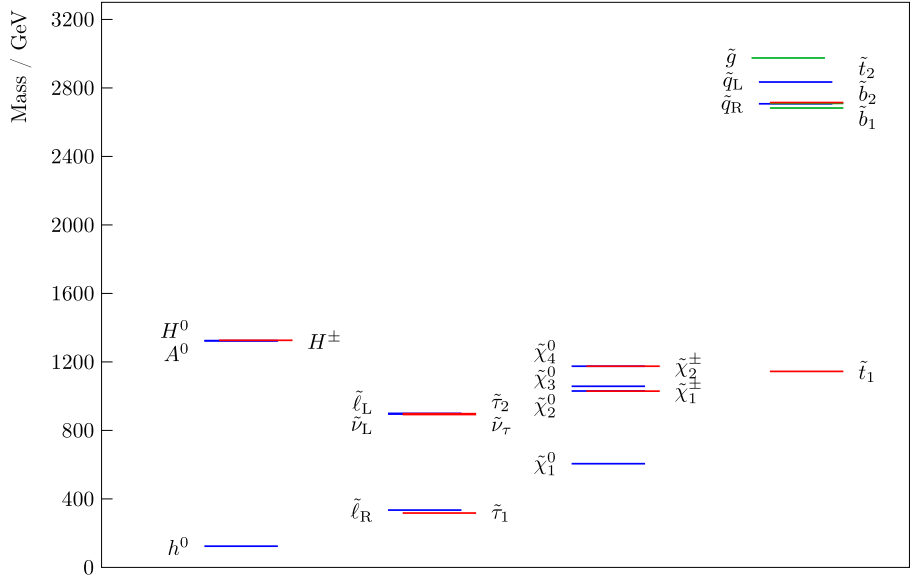}
\caption{Spectrum for the point marked with the dot in figure \ref{fig:nmfv}. The Higgs mass is $m_h=124$ GeV, with $A_t=-2.9$ TeV. $\tilde{\tau}_1$ is $17$ GeV lighter than the right handed sleptons. The Higgsino mass is $\mu=1.05$ TeV.  Fine tuning is $\sim 1/400$.}
\label{fig:NMFVspec}
\end{center}
\end{figure}

A typical spectrum for the model is presented in figure \ref{fig:NMFVspec}, which corresponds to the black dot indicated in the two different planes presented in figure \ref{fig:nmfv}.\footnote{We choose our benchmark point here and in the next subsection to have $m_h=124$~GeV in order to account optimistically for the theory uncertainty on the Higgs mass calculation.} In general, the spectrum across the parameter space of our model is basically that of MGM with $N_{mess}=3$ (gaugino unification, colored sparticles heavier than electroweak sparticles, right-handed stau NLSP, etc.). There are, however, two key differences. First, in order to counteract the large negative contribution  (\ref{eq:littleAmu}) to the right-handed stop, the MGM scale $\frac{F_{X}}{M}$ is considerably larger than would otherwise be the case. This results in the other colored sparticles being essentially decoupled. It also results in a higher gravitino mass, which explains \cite{Ambrosanio:1997bq} why slepton co-NLSPs do not occur in figure \ref{fig:nmfv}. Second, the right-handed sleptons are a bit lighter than in MGM due to the effects of running induced by the split stops. Amusingly, this effect of running means that the stau is the NLSP even for lower $N_{mess}$,  unlike in MGM, where lowering $N_{mess}$ leads to bino NLSP.\footnote{Note that if we exchange the roles of $\bar{u}_3$ and $Q_3$ in (\ref{exampleU}), a negative soft mass for $Q_3$ would be induced instead, leading to a heavier $\tilde{\tau}_1$ through running. In this case, it could be possible to have a bino NLSP even for $N_m>1$.}

Due to the split spectrum, the largest sparticle pair production cross sections at LHC correspond to $\tilde{t}_1$ and the right-handed sleptons. Pair production of stops leads to a decay chain with jets, leptons and missing energy. When right handed sleptons are directly pair produced, the decay chain will include relatively soft leptons (due to the moderate splitting of the right handed sleptons and the stau), taus and missing energy. Of course the direct pair production of staus will lead to taus and missing energy.

Of the above signatures, the most spectacular one is given by the decay of pair produced stops, which can contain two jets, 4 leptons (from the decay of the bino to RH sleptons and RH sleptons to stau), and two $\tau$ jets plus missing energy. A search with a similar topology was carried out in \cite{Aad:2014mra}, where a limit on the total strong production cross section of $\sim1$ fb was obtained. This limit can be used to set an approximate bound on our parameter space, by comparing with our model's tree level total strong production cross section, which we obtain using {\sc MadGraph} \cite{Alwall:2014hca}. This leads to excluding stops roughly below $800$ GeV in the parameter space presented in figure \ref{fig:nmfv}, which corresponds to staus heavier than $150$ GeV. 

The spectrum presented in figure \ref{fig:NMFVspec} is inaccessible to the LHC run at $8$ TeV, but it will become accessible at $14$ TeV. The total SUSY cross section of such point at the $14$ TeV LHC is $~8$ fb, while the total tree level colored production cross section is $~2$ fb. Relevant searches will be the updated versions of multilepton or GMSB-inspired searches as  \cite{Chatrchyan:2014aea} and \cite{Aad:2014mra}.

\section{An MFV model}
\label{sec:MFV}

Next we will turn to the MFV model (\ref{exampleH}). Apart from the issues of UV completions to be discussed in the next section, this model is slightly more complicated than the non-MFV model because here we would like to generate the MSSM up-type Yukawas and the $A$-terms from the same operator. To achieve this, it is necessary to turn on a lowest component vev for $X'$, which implies that one must re-diagonalize the messenger mass matrix prior to integrating out the messengers. For later convenience, we will redefine $X'$ so that its lowest component vev is separated out and denoted by  $X_0'$. Then (\ref{exampleH}) becomes
\begin{equation}\label{minimal MFV}
W = (X'_0+X')H_u \tilde \phi  + \lambda_u^{ij} \phi Q_i \bar{u}_j + M\phi \tilde \phi \end{equation}
with $\langle X'\rangle=F_{X'} \theta^2$. The form of (\ref{minimal MFV}) is the most general allowed by a $\mathbb{Z}_3$ symmetry, as detailed in table \ref{tab:Z3}, which also allows for a $\mu$-term and down type Yukawas, 
 \begin{equation}
 \label{downtypeYuk}
\delta W = \mu' H_u  H_d + \lambda_d^{ij} H_d Q_i \bar{d}_j + \lambda_e^{ij} H_d L_i \bar{e}_j 
\end{equation}
We will not discuss the down sector Yukawas any further. 

\begin{table}[t]
\begin{center}
\begin{tabular}{|c|cccccc|}\hline
&$X'$&$Q,\bar{u},\bar{d},L,\bar e$ &${H}_u$& $H_d$&  $\phi$ &$\tilde{\phi}$  \\\hline
$\mathbb{Z}_3$&1/3&2/3&1/3&2/3&2/3&1/3 \\\hline
\end{tabular}
\end{center}
\caption{Charge assignments securing (\ref{minimal MFV}) and (\ref{downtypeYuk}).}
\label{tab:Z3}
\end{table}

After diagonalizing the mass matrix and integrating out the heavy messenger states, we are left with the supersymmetric effective action:
\bea\label{MFVeffWK}
W_{eff}&\supset y_u^{ij}\left(1+\cot\theta_H\cos\theta_H\frac{X'}{M'}\right) H_u Q_i \bar{u}_j +\mu\left(1+\sin \theta_H \frac{X'}{M'}\right) H_u H_d \\ 
K_{eff}&\supset \frac{\cos^2 \theta_H}{M'^2}X^{'\dagger} X' H_u^\dagger H_u+ \frac{\cot^2 \theta_H}{M'^2}y_u^{il} {y_u^{jk}}^{*} Q_i^\dagger u_l^\dagger Q_j u_k   \eea
where
\bea 
& M'=\sqrt{X^{'2}_0+M^2},\qquad \sin \theta_H=\frac{X'_0}{M'},\qquad y_u^{ij}=-\lambda_u^{ij}\sin\theta_H,  \qquad \mu= \mu'\cos\theta_H
  \label{eq:rotationMFV}
\eea
and we have everywhere expanded in $\mu'\ll M,X_0$, keeping only the lowest nonzero order. 
In (\ref{MFVeffWK}), the first term in the effective superpotential  leads to an $A$-term proportional to the up-type Yukawas. The second term in the effective K\"ahler potential  is an MFV interaction suppressed by the messenger scale, so it is safe from flavor constraints \cite{Isidori:2010kg}. Meanwhile, the first term in $K_{eff}$ represents a contribution to the soft mass of $H_u$:\footnote{The second term in the effective superpotential (\ref{MFVeffWK}) gives rise to $B_\mu=  \mu A_t\tan^2 \theta_H$ at the messenger scale. While this is parametrically of the right size for EWSB, it has the incorrect sign to lead to the large $\tan\beta$ EWSB condition $B_\mu\approx 0$ at the weak scale. Thus a more complete model that also aspires to explain the origin of $\mu$ and $B_\mu$ must include additional contributions to these parameters.}
\begin{equation}
\delta m^2_{H_u} 
= - |A_t|^2\tan^2 \theta_H
\label{eq:MFVsoftmass}
\end{equation}
This is a manifestation of the little $A/m_H^2$ problem.
Note that this contribution is negative, so it is not dangerous for electroweak symmetry breaking, unlike what was found in the K\"ahler potential models \cite{Craig:2012xp}. However, if $\tan \theta_H\gtrsim 1$ it still represents a major contribution to fine-tuning. Taking $\tan \theta_H\ll 1$ would alleviate this fine-tuning problem, but at the cost of enlarging the underlying coupling $\lambda_u^{33}$  according to (\ref{eq:rotationMFV}). This leads to a Landau pole at low scales and a UV completion becomes necessary. Such a UV completion is the subject of section \ref{sec:fullycomposite}, in which we use Seiberg duality \cite{Seiberg:1994pq, Seiberg:1994bz} to realize the large coupling $\lambda^{33}_u$.

As in the previous section, to generate the rest of the soft masses we specialize to the case of MGM.  The parameter space is essentially the same as before, namely the MGM sector is described by $N_m$, $\tan\beta$, $M$ and $\frac{F_{X}}{M}$, while our effective theory contains $\frac{F_{X'}}{M}$ which sets the scale for the tree level contribution to $A_t$, and a coupling $\lambda_u^{33}$.  Again, we consider $\mu$ and $B_{\mu}$ to be determined by the EWSB conditions. We fix most of the parameters to the same values as before -- $N_m=3$, $\tan\beta=20$ and $M=250$~TeV -- for essentially the same reasons. 
 Finally, we consider two values for $\lambda_u^{33}$: $\lambda_u^{33}=1$ is chosen to illustrate the perturbative case, while $\lambda_u^{33}=3$ is studied  since it has a beneficial effect on decreasing tuning. With these choices, the parameter space of our model reduces to $(\frac{F_{X'}}{M}, \frac{F_{X}}{M})$, which we can trade for the IR values of the $A$-term $A_t$ and the gluino mass $M_{\tilde{g}}$.

\begin{figure}[t!] 
\centering
\begin{subfigure}{0.5 \textwidth}
\psfrag{Y}[Bc]{\small $M_{\tilde{g}}$ \normalsize}
\psfrag{X}[lc]{ \small $A_t$  \normalsize}
	\centering
	\includegraphics[width=7cm]{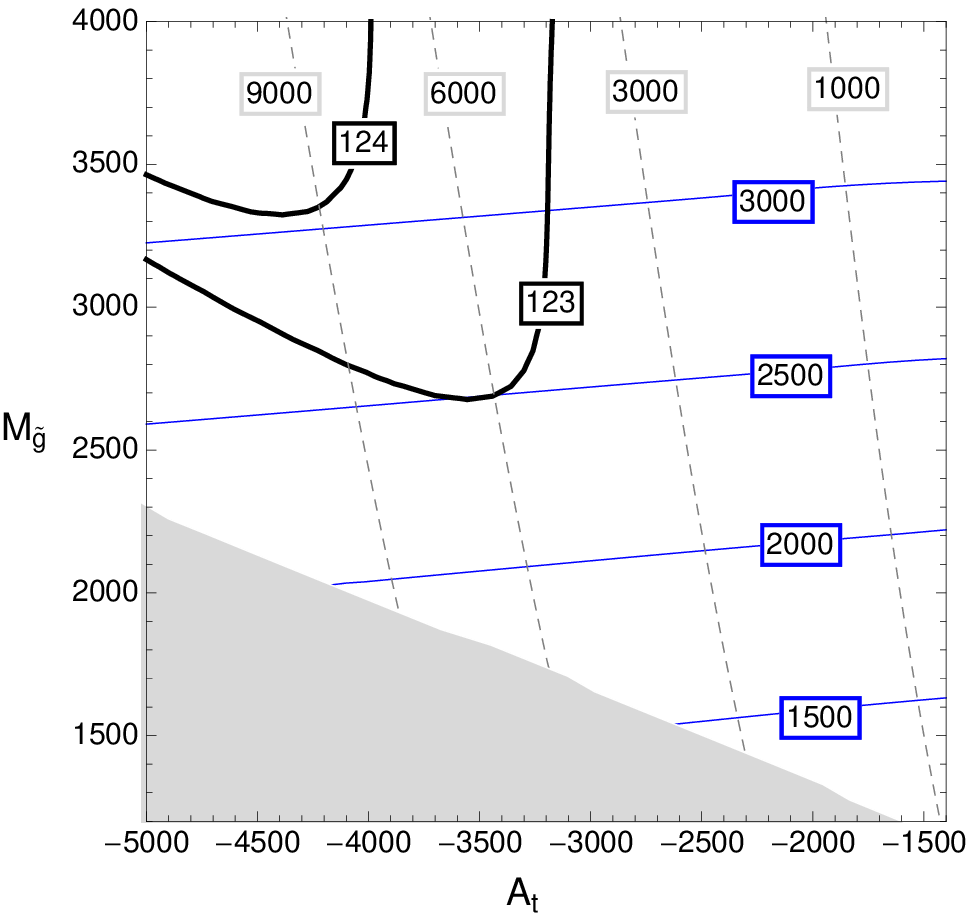}
	\caption{$\lambda_u^{33}=1$}
	\label{fig:k1}
\end{subfigure}%
\begin{subfigure}{0.5 \textwidth}
\psfrag{Y}[Bc]{\small $M_{\tilde{g}}$ \normalsize}
\psfrag{X}[lc]{ \small $A_t$  \normalsize}
	\centering
	\includegraphics[width=7cm]{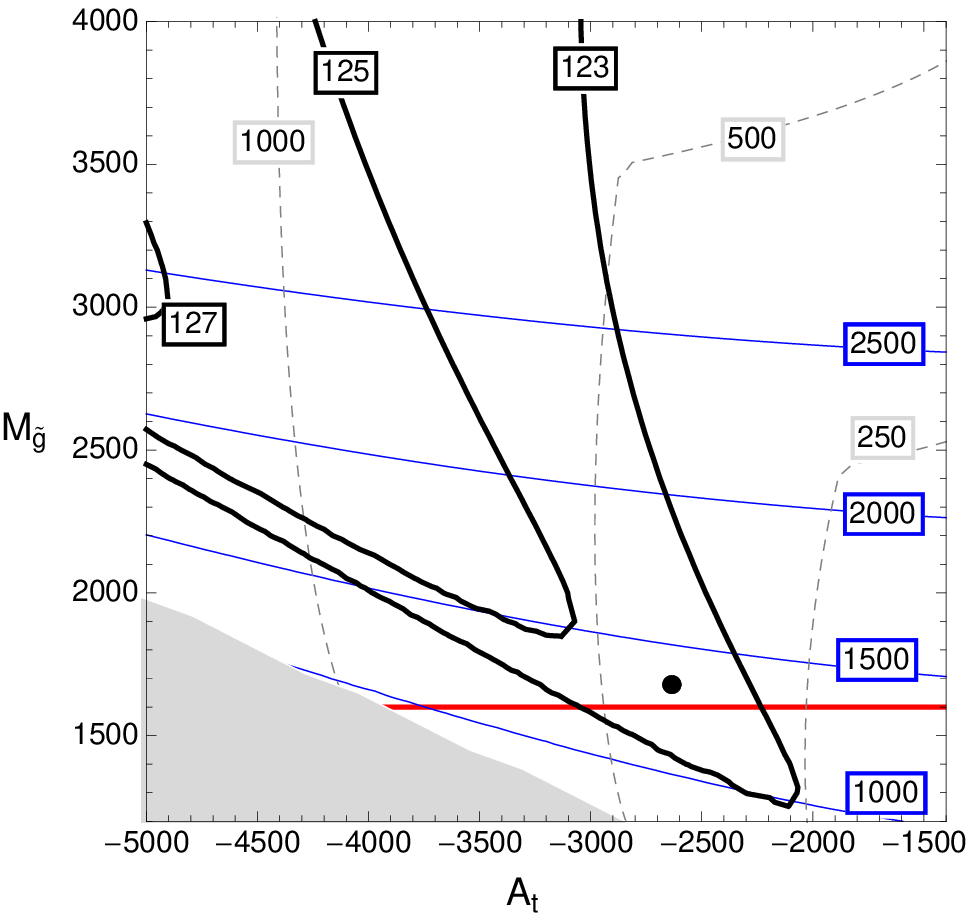}
	\caption{$\lambda_u^{33}=3$}
	\label{fig:k3}
\end{subfigure}
\caption{Contours of the Higgs mass (black), geometric mean of the stop masses (blue) and tuning (dashed), for two choices of $\lambda_u^{33}$ with $N_m=3$, $\tan\beta=20$, $M=250$ TeV. Different Higgs mass contours are presented to account for the uncertainty in the theoretical Higgs mass calculation. The shaded region corresponds to tachyonic stops/staus. The dot on the figure on the right corresponds to the point in parameter space with the spectrum presented in figure \ref{fig:largekappa}. The parameter space below the red line on the same figure is excluded by \cite{Aad:2014mra}. All quantities are evaluated at $M_{SUSY}$.}\label{fig:MFVhiggs}
\end{figure}

In figure \ref{fig:MFVhiggs} we show contours of the Higgs mass, tuning and $M_{SUSY}$ in the $(M_{\tilde{g}}, A_t)$ plane for the two choices of $\lambda_u^{33}$. In both figures \ref{fig:k1} and \ref{fig:k3} a large Higgs mass can be achieved with moderate values of $M_{SUSY}$ thanks to the large $A$-terms. In figure \ref{fig:k1} however, the $\mu$-term is very large and induces sizable negative contributions to $m_h$ through the stau and sbottom sectors. This implies that a higher $M_{SUSY}$ is needed to obtain the correct Higgs mass. (see e.g. \cite{Carena:1995wu}.) The main source of tuning can be either the large induced Higgs soft mass from  (\ref{eq:MFVsoftmass}) or, for large $M_{SUSY}$, the running effect. We immediately see from figure \ref{fig:k1} that the first of these sources represents a serious tuning problem for $\lambda_u^{33}=1$, in which case for a $125$ GeV Higgs we obtain a typical tuning of  $\sim 10^{-4}$. In figure \ref{fig:k3} we see the beneficial effect of considering a larger value for  $\lambda_u^{33}$. This choice suppresses the fine tuning induced by  (\ref{eq:MFVsoftmass}), in such a way that a $125$ GeV Higgs can be achieved while keeping tuning to the one part in $\sim 500$ level. 

\begin{figure}[t]
\begin{center}
\includegraphics[width=12cm]{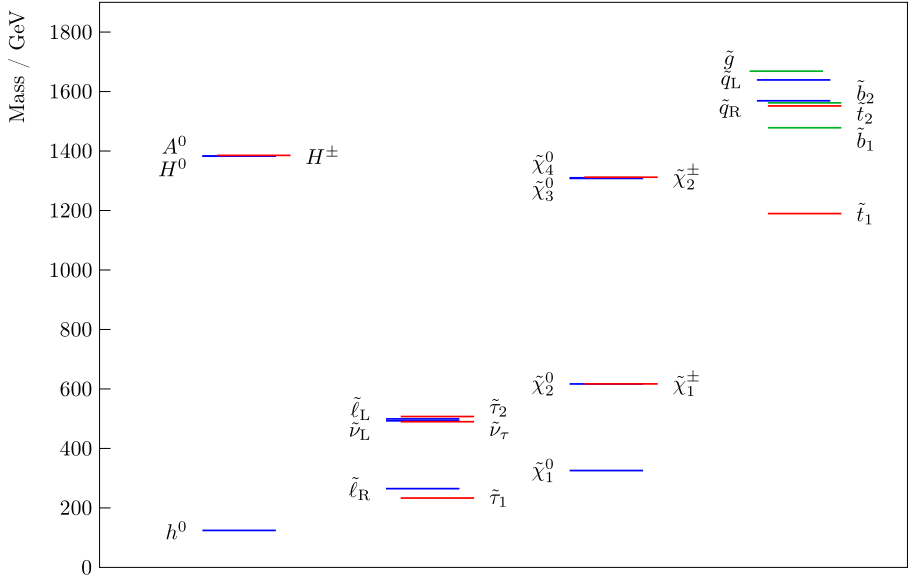}
\caption{Spectrum for the point shown in figure \ref{fig:k3}. The Higgs mass is $m_h=124$ GeV, with $A_t=-2.7$ TeV. $\tilde{\tau}_1$ is $32$ GeV lighter than the right handed sleptons. The Higgsino mass is $\mu=1.3$ TeV. Fine tuning is $\sim 1/400$.}
\label{fig:largekappa}
\end{center}
\end{figure}

In figure \ref{fig:largekappa} we present a typical spectrum for the model with $\lambda_u^{33}=3$, which corresponds to the black dot in figure \ref{fig:k3}. This model is even more similar to MGM with stau NLSP than the one presented in the previous subsection, since there is no negative contribution to the right-handed stop to counteract. The only difference now with MGM is the large $A$-term, which has a minor effect on the rest of the spectrum primarily through the RG. The MGM collider signatures here are potentially spectacular. If colored superpartners are accessible to collider experiments they will lead to a long decay chain including jets, leptons and missing energy. As in our non-MFV model, searches that look for jets, tau final states and large missing energy can be sensitive to this spectrum when the strong production is accessible. In particular ATLAS search \cite{Aad:2014mra} analyses a similar spectrum and their results apply directly to our case, setting strong bounds on parts of the parameter space.  For $\tan\beta=20$, gluinos of up to $1.6$ TeV are excluded, which corresponds to a total strong production cross section of $\sim1.5$ fb at tree level   \cite{Alwall:2014hca}. 

Multilepton searches could also be a leading probe of this model, especially when the colored sparticles are too heavy to be produced. The stau NLSP scenario considered in \cite{Chatrchyan:2014aea} can be sensitive to our case, but since in our spectrum $\tilde{m}_{e_R}-\tilde{m}_{\tau_1}\sim 20$ GeV and $150 \text{ GeV}<\tilde{m}_{\tau_1}$, the obtained bounds are not currently relevant for us. However, updates of these searches in Run II of the LHC can be very interesting for our models.

\section{A composite model from Seiberg duality}
\label{sec:fullycomposite}

As discussed in the previous section, the little $A/m_H^2$ problem in the MFV model (\ref{eq:MFVsoftmass}) necessitates a large value for $\lambda_u^{33}$, and the theory has a Landau pole at a low scale. One way to explain physics above the Landau pole is to build composite models that naturally provide $|\lambda_{u}^{33}|\gg 1$ due to the underlying strong interactions. In general, characterizing such a strongly coupled UV completion is challenging at best, however in the context of supersymmetric gauge theories we can make use of Seiberg duality \cite{Seiberg:1994pq, Seiberg:1994bz}. We embed the model of section \ref{sec:MFV} in the magnetic side of the duality, where the fields $Q_3, \bar{u}_3$ and $\phi$ will be composite degrees of freedom. Since it is conceptually simpler, we first discuss the electric side of the duality. In a second stage we discuss the mapping to the composite degrees of freedom on the magnetic side, and we complete the model by adding in a number of spectator fields.

\subsection{Electric theory}
\label{subsec:electric}

The electric theory is defined by SQCD with $N_c=2$ colors and $N_f=3$ flavors. Since the fundamental of the electric gauge group $SU(2)_E$ is pseudo-real, this theory is invariant under an $SU(6)$ global symmetry. It is therefore convenient to parametrize its degrees of freedom with a single matter field $q^i_a$ in the fundamental of $SU(2)_E$ and $SU(6)$. The standard model gauge group can be embedded in the global symmetry as follows
\begin{equation}\label{embedding}
SU(6)\supset SU(5)\supset SU(3)_c \times SU(2)_L \times U(1)_Y
\end{equation}
With this matter content, the global symmetry is anomalous. In section \ref{subsec:spectators} we will introduce some spectator fields to cancel the gauge anomalies and give vector-like masses to some exotics. Note that because the global symmetry contains $SU(5)$, grand unification is manifest in this model from the outset. Concretely, the fundamental of $SU(6)$ trivially decomposes as 
\begin{equation}
\mathbf{6}=\mathbf{5}\oplus \mathbf{1}
\end{equation}  
where the $\mathbf{5}$ further decomposes into standard model representations in the conventional way. The quantum numbers of $q_a^i$ are summarized in table \ref{tab:electricmattercontent}. 

In addition to hypercharge $U(1)_Y$, the breaking pattern in (\ref{embedding}) allows for an additional global symmetry which we will denote by $U(1)_G$. As will be seen in section \ref{subsec:magnetic}, it is necessary to consider the MSSM baryon number to be part of the global symmetries for proton stability. It will also be seen that baryon number has a unique embedding in $U(1)_G$ and $U(1)_Y$ given by:
\begin{equation}\label{baryongen}
B=\frac{4}{5} Y+\frac{1}{10} G \quad\quad \mathrm{with} \quad\quad \begin{array}{l}Y=\text{diag}(-\frac{1}{3},-\frac{1}{3},-\frac{1}{3},\frac{1}{2},\frac{1}{2},0)\\G=\text{diag}(1,1,1,1,1,-5)\end{array}
\end{equation}
Note that both the electric and magnetic theories have a $\mathbb{Z}_{N_f}$ discrete symmetry that is leftover from the anomalous global $U(1)$ symmetry. As we will discuss in the next subsection, we will identify this $\mathbb{Z}_3$ with the one of table \ref{tab:Z3}.

\begin{table}[t]\centering
\begin{tabular}{|c|c|c|ccc|c|c|}\hline
GUT&field&$SU(2)_E$& $SU(3)_c$ & $SU(2)_L$ & $U(1)_Y$&$\mathbb{Z}_3$&$B$\\\hline
\multirow{2}{*}{$\mathbf{5}$}&$q_c$ & $\square$ & $\square $&1&$-\frac{1}{3}$&$\frac{1}{3}$&$-\frac{1}{6}$\\
&$q_L$ & $\square$ & $1$& $\square$&$\frac{1}{2}$&$\frac{1}{3}$&$\frac{1}{2}$\\\hdashline
$\mathbf{1}$&$q_S$ & $\square $ & $1 $&1&$0$&$\frac{1}{3}$&$-\frac{1}{2}$\\\hline
\end{tabular}
\caption{Matter content of the electric theory. $q=q_c\oplus q_L\oplus q_S$ form a fundamental of the $SU(6)$ global symmetry. \label{tab:electricmattercontent} }
\end{table}

\subsection{Magnetic theory}
\label{subsec:magnetic}

This theory s-confines in the IR and has a weakly-coupled magnetic dual description in terms of the mesons and baryons of the electric theory as described in table \ref{tab:magneticmattercontent}. These gauge invariants $q^iq^j$ transform as the antisymmetric tensor $\mathbf{15}_A$ of the global $SU(6)$. Under $SU(5)$ this decomposes as
\begin{equation}
\mathbf{15}_A= \mathbf{10}_A\oplus \mathbf{5}.
\end{equation}

\begin{table}[!t]\centering
\begin{tabular}{|c|c|ccc|c|c|c|}\hline
GUT &field& $SU(3)_c$ & $SU(2)_L$ & $U(1)_Y$&composite&$\mathbb{Z}_3$&$B$\\\hline
\multirow{3}{*}{$\mathbf{10}$}&$\color{black}{Q_3}$ & $\square$&$\square$&$1/6$&$q_c q_L$&$2/3$&1/3\\
&$\color{black}{\overline u_3}$ & $\overline{\square}$&$1$&$-2/3$&$q_c q_c$&$2/3$&-1/3\\
&$E'$& $1$&$1$&$1$&$q_L q_L$&$2/3$&1\\\hdashline

\multirow{2}{*}{$\mathbf{5}$}&$\phi$ & $1$&$\square$&$1/2$&$ q_L q_S$&$2/3$&0\\
&$d'$ & $\square$&$1$&$-1/3$&$q_c q_S$&$2/3$&-2/3\\\hline
\end{tabular}
\caption{Matter content of the magnetic side of the duality. All fields fill out complete GUT multiplets. Since $E'$ carries baryon number, it cannot be identified with a right handed lepton. \label{tab:magneticmattercontent}}
\end{table}

The resulting $SU(5)$ representations allow us to identify $Q_3$, $\bar{u}_3$ and $\phi$ with composite degrees of freedom. Note that the baryon numbers of $Q_3$ and $\bar{u}_3$ uniquely determined the coefficients of $U(1)_Y$ and $U(1)_G$ in (\ref{baryongen}). The rest of the composite fields are $E'$ and $d'$, of which $E'$ has the same gauge quantum numbers as right handed leptons, but non-zero baryon number.

The confining electric gauge group dynamically generates a superpotential in the magnetic dual, given by
\bea\label{pfaff}
W_{\mathrm{mag}} &= \frac{1}{\Lambda^3}\mathrm{Pf} (q^i q^j)\\
&= \kappa ( \phi {Q}_3 {\bar{u}}_3 - {Q}_3{Q}_3 d'+d' \bar{u}_3 E')\label{eq:magpotential}
\eea
where $\mathrm{Pf}$ is the Pfaffian of the antisymmetric matrix $q^i q^j$, and we used the mapping to the magnetic theory in the second line.
The coupling $\kappa$ descends from the strong dynamics in the electric theory and can be large (for concreteness we assumed $\kappa \sim 3$ in section \ref{sec:MFV}). From the last two operators in (\ref{eq:magpotential}) it should also be clear that rapid, dimension 6 proton decay would be introduced if one were to identify $E'$ with one of the MSSM leptons. The $B$ and $\mathbb{Z}_3$ charges for the composite fields are fixed by those of the electric quarks in table \ref{tab:electricmattercontent}.

\subsection{Complete model with spectators}
\label{subsec:spectators}
Let us now weakly gauge a $SU(3)_c \times SU(2)_L \times U(1)_Y$ subgroup of the global symmetry. To cancel anomalies, fill out complete GUT multiplets, and match the field content of the magnetic theory to the model of section \ref{sec:MFV}, we add a number of fundamental fields, which are all spectators as far as the Seiberg duality is concerned. Among these spectators are all three $\bar d$, $L$ and $\bar e$ generations of the MSSM, as well as the first two generations of the $Q$ and $\bar u$ sectors. Finally, the $H_u$ and $H_d$ are spectators as well, but do not come in complete GUT multiplets. This is nothing other than the usual doublet-triplet splitting problem in models with grand unification.  The spectators and their quantum numbers are introduced in table \ref{tab:spectatormattercontent}. Aside from the usual baryon number, we also assign the $\mathbb{Z}_3$ charges for the spectator fields such that the symmetry in table \ref{tab:Z3} is realized. In addition to the fields we introduced so far, one may choose to add up to three pairs of conventional, $\mathbf{5}$-$\mathbf{\overline{5}}$ gauge mediation messengers without spoiling perturbative gauge coupling unification.\footnote{We hereby assume that any uncalculable threshold corrections at the compositeness scale are negligible.}

\begin{table}[!t]\centering
\begin{tabular}{|c|c|ccc|c|c|}\hline
GUT &field& $SU(3)_c$ & $SU(2)_L$ & $U(1)_Y$&$\mathbb{Z}_3$&$B$\\\hline
\multirow{2}{*}{$\mathbf{\overline 5}$}&$\tilde{\phi}$ & $1$&$\square$&$-1/2$& $1/3$&0\\
&$\overline d'$ & $\overline{\square}$&$1$&$1/3$&$1/3$&2/3\\\hdashline

\multirow{2}{*}{$\mathbf{\overline 5}$}&$\color{black}{L_3}$ & $1$&$\square$&$-1/2$&$2/3$&0\\
&$\color{black}{\overline{d}_3}$ & $\overline{\square}$&$1$&$1/3$&$2/3$&-1/3\\\hdashline

\multirow{3}{*}{$\mathbf{10}$}&$Q'$ & $\square$&$\square$&$1/6$&$1/3$&1/3\\
&$\overline U'$ & $\overline{\square}$&$1$&$-2/3$&$1/3$&-1/3\\
&$\color{black}{\bar{e}_3} $& $1$&$1$&$1$&$2/3$&0\\\hdashline

\multirow{3}{*}{$\mathbf{\overline{10}}$}&$\overline Q'$ & $\overline\square$&$\square$&$-1/6$&$2/3$&-1/3\\
&$ U'$ & ${\square}$&$1$&$2/3$&$2/3$&1/3\\
&$\overline E' $& $1$&$1$&$-1$&$1/3$&-1\\\hdashline

\multirow{1}{*}{$$}&$H_u$ & $1$&$\square$&$1/2$& $1/3$&0\\\hdashline

\multirow{1}{*}{$$}&$H_d$ & $1$&$\square$&$-1/2$& $2/3$&0\\

\hline

\end{tabular}
\caption{Spectators of the Seiberg duality required to cancel anomalies and fill out complete GUT multiplets. Primed fields have heavy vector-like masses and are integrated out at the duality scale. The first two generations are also spectators but are not shown here for simplicity. \label{tab:spectatormattercontent}}
\end{table}

All the non-MSSM fields have vector-like masses. Some arise from Yukawa interactions in the electric theory, while others are mass terms: 
\bea\label{eq:electricmass}
W_{\mathrm{elec}}&\supset y_{d'} q_c q_S \overline{d}'+y_{E'} q_L q_L \overline{E}' + M_{Q'}Q' \bar Q' + M_{U'}U'\bar U'\\
 &\qquad \rightarrow W_{\mathrm{mag}}\supset y_{d'} \Lambda d' \overline{d}'+y_{E'} \Lambda E' \overline{E}'+ M_{Q'}Q' \bar Q' + M_{U'}U'\bar U'
\eea
Those that are Yukawas in the electric theory are naturally of the same size as the compositeness scale $\Lambda$, and so for unification we must also take $M_{Q'}\sim M_{U'}\sim \Lambda$.

We can see that it is possible to reproduce the model in (\ref{minimal MFV}) by adding interactions between spectators and the composites and between spectators themselves if we allow the following interactions
\bea\label{eq:spectpotential}
\delta W &=(X_0'+X') H_u \tilde{\phi}  + \tilde{\lambda}_{u}^{ij} \phi {Q}_i {\bar{u}}_j + M \phi \tilde{\phi}
\eea
where $i,j$ identify quark fields in the gauge eigenbasis. To avoid clutter, we suppressed the mass terms that are introduced in (\ref{eq:electricmass}), as well as the $\mu$-term and the down and lepton Yukawas. This superpotential is generic if we impose the $\mathbb{Z}_3$ symmetry of tables \ref{tab:magneticmattercontent} and \ref{tab:spectatormattercontent}.  

As noted earlier, the first and second generations of the MSSM matter fields are all elementary and spectators as far as the Seiberg duality is concerned. Since $\phi$ is a composite operator in the electric theory, all up-type Yukawa couplings (other than the top Yukawa) must arise from irrelevant operators in the electric theory. (Recall that the ${\Bbb Z}_3$ symmetry of table \ref{tab:Z3} forbids the usual up-type Yukawa couplings $H_u Q \bar u$.) For instance
\bea\label{eq:irrelev1}
\frac{1}{\Lambda_{UV}^2}(q_L q_S)(q_c q_L)\bar u_2\quad&\rightarrow \quad\frac{\Lambda^2}{\Lambda_{UV}^2}\phi Q_3\bar u_2\\
\frac{1}{\Lambda_{UV}}(q_L q_S)Q_2\bar u_2\quad&\rightarrow\quad \frac{\Lambda}{\Lambda_{UV}}\phi Q_2\bar u_2\label{eq:irrelev2}
\eea
where $\Lambda_{UV}$ is a cut-off scale of the electric theory. In the notation of section  \ref{sec:MFV} this yields:
\bea
\label{eq:yukawa}
\lambda_u^{ij}&=  \kappa \delta^{i3}\delta^{3j}+\tilde \lambda_u^{ij}\sim\left(\!\!\begin{array}{ccc}0&0&0\\0&0&0\\0&0&\kappa\end{array}\!\!\right)+
 \left(\!\!\begin{array}{ccc}\epsilon&\epsilon&\epsilon^2\\\epsilon&\epsilon&\epsilon^2\\\epsilon^2&\epsilon^2&\epsilon^3\end{array}\!\!\right)
\eea
with  $\epsilon\sim \Lambda/\Lambda_{UV}\ll 1$. The composite sector therefore naturally provides a partial explanation of the texture of the up-type Yukawa matrix. 
Since $Q_3$ is a composite degree of freedom, it also predicts $\epsilon \sim y_b \sim 0.1$, but the rest of the hierarchies in $y_d$ and $y_\ell$ are not explained. 

Upon integrating out the messenger fields, the analysis further reduces to what was presented in section \ref{sec:MFV}. There is one exception, in the sense that the model is no longer manifestly MFV since the third generation was given a special treatment. In particular a non-MFV dimension six operator is generated in the K\"ahler potential from integrating out $d'$ in (\ref{eq:magpotential})
\begin{equation}
\delta K_{\mathrm{eff}}\sim {1\over \Lambda^2}(Q_3 Q_3)^\dagger (Q_3 Q_3)\sim  {1\over \Lambda^2} (u_3 d_3)^\dagger (u_3 d_3).
\end{equation}
By rotating $Q_3$ to the mass eigenbasis, this operator can in principle couple quarks of different generations. However note that this operator does not introduce any new CP phase into the model and it does not contribute to FCNC processes at tree level. Moreover it is suppressed by the duality scale that is above the messenger scale $\gtrsim 100$~TeV. The effects in the first two generation quarks are further suppressed by powers of $\epsilon$ coming from (\ref{eq:yukawa}). For instance, the operator contributing to $K$-$\bar K$ mixing receives an additional suppression of $\sim \epsilon^8$. Therefore we conclude that it is consistent with the bounds from flavor observables \cite{Isidori:2010kg}. 

\section{Conclusions}

In this paper, we presented a new mechanism to generate large $A$-terms through tree-level superpotential operators. We provided explicit examples of both MFV and non-MFV models. In contrast to the conventional setups with one-loop $A$-terms through K\"ahler potential operators, our tree-level mechanism does not induce any dangerously large soft masses and is therefore manifestly free from the $A/m^2$ problem. 
 Generically, a soft mass of the same order as the $A$-term is nevertheless still generated. For the non-MFV example this contribution greatly increases the splitting between the stop mass eigenstates, but otherwise does not significantly impact the phenomenology or the fine tuning. For the MFV case, the soft mass could potentially lead to disastrous levels of fine tuning, but it can be brought under control by the existence of strong dynamics near the messenger scale. We provide an example of such a composite sector which has a description in terms of Seiberg duality and which explicitly allows for  gauge coupling unification.

Some potential future directions suggested by this work include:
\begin{itemize}

\item For concreteness, we focused on an MGM setup as a first example, but we emphasize that tree-level $A$-terms are merely a module that can be added to any mechanism for mediating SUSY breaking. In particular, it would be interesting to study whether the mechanism can naturally be embedded in more realistic models of dynamical supersymmetry breaking. In addition one could generalize $X'$ beyond the spurion limit, and study the effects of its dynamics on the phenomenology. 

\item In the non-MFV case it may be interesting to embed the tree-level $A$-term into a full fledged theory of flavor. 

\item In the MFV case, we saw that the $A$-term module generated a contribution to $B_\mu$ which unfortunately was of the wrong sign for EWSB. An interesting opportunity here would be to construct a complete model that produces both tree-level $A$-terms and $B_\mu$, perhaps along the lines of  the models constructed in \cite{Komargodski:2008ax}. 

\item Finally, the emergence of large $A$-terms from a composite sector in the MFV case may open a new avenue towards constructing a realistic model where large $A$-terms are generated at the TeV scale, hence further reducing the fine-tuning.

\end{itemize}

\section*{Acknowledgments}
We thank David E.~Kaplan for discussions that originally inspired this work. We also thank Nathaniel Craig, Jared Evans, Diego Redigolo, Arun Thalapillil and Scott Thomas for useful discussions. We are grateful to Ben Allanach for advice regarding \texttt{SOFTSUSY 3.5.1} and to Diego Redigolo for comments on the manuscript. The work of A.B.\ and D.E.\ is supported by DOE-ARRA-SC0003883 and DOE-SC0010008.  D.E. is further supported by CONICYT Becas Chile and the Fulbright Program. The work of D.S.\ is supported by a DOE Early Career Award and a Sloan Foundation Fellowship. This manuscript has been authored by an author (SK) at Lawrence Berkeley National Laboratory under Contract No. DE-AC02-05CH11231 with the U.S. Department of Energy. The U.S. Government retains, and the publisher, by accepting the article for publication, acknowledges, that the U.S. Government retains a non-exclusive, paid-up, irrevocable, world-wide license to publish or reproduce the published form of this manuscript, or allow others to do so, for U.S. Government purposes.

\appendix

\section{The little $A/m^2$ problem for arbitrary couplings}
\label{app:generalization}

In sections \ref{sec:NMFV models} and \ref{sec:MFV} we concluded that the little $A/m^2$ tuning problem is most serious when a soft mass for the Higgs field is generated. In this appendix we show that this little $A/m^2$ problem is generic for our class of models: it cannot be avoided by increasing the messenger number or considering a more general renormalizable superpotential. 

Consider the most general renormalizable superpotential coupling the fields $H_u, Q, \bar{u}$ with $n$ pairs of messengers $\phi_k,\tilde{\phi}_k$ ($k=1\dots n$) with the quantum numbers of $H_u$ and its hermitian conjugate 
\bea
W=M_k\phi_k\tilde{\phi}_k+X_k \tilde{\phi}_k H_u+y_t H_u Q \bar{u}+\lambda_k \phi_k  Q\bar{u} + \dots
 \label{eq:superpotential}
  \eea
where we sum over repeated indices. Here, differently from section \ref{sec:MFV}, we work in a basis in which the supersymmetric mass matrix has already been diagonalized, so $X_k$ have only F-term vevs. The rest of the interactions included in $\dots$ do not matter to derive the induced $A$-term and soft mass at lowest order, so we neglect them in what follows. Integrating out the messengers in the small SUSY breaking regime $F/M^2 \ll 1$ we get the low energy superpotential and K\"ahler potential
\begin{align}
 W=y_t H_u Q \bar{u} - \lambda_k \frac{X_k }{M_k} H_u Q \bar{u}  \quad , \quad K=\left(1+\left(\frac{X_k  }{M_k}\right)^{\dagger}\left(\frac{X_k }{M_k}\right)\right) H_u^\dagger H_u+ \dots
\label{eq:lowenergyW}
\end{align}
so the $A$-term and induced soft mass are
\begin{equation}
y_tA_t=- \frac{\lambda_kX_k}{M_k} \quad , \quad \delta m^2_{H_u}=-\left(\frac{X_k}{M_k}\right)^{\dagger}\left(\frac{X_k}{M_k}\right) 
\label{eq:inducedmass}
\end{equation}
To avoid the little $A/m^2$ problem, we need to maximize the ratio of the $A$-term over the soft mass. In particular we are interested in knowing if in doing this, the theory remains perturbative, or if it does not, when does it become strongly coupled. To address this question, note that there is a linear combination of messengers that couples to the light fields with a Yukawa with magnitude given by
\begin{equation}
 \abs{\lambda}=\sqrt{\sum_k \abs{\lambda_k}^2}
\end{equation}
so that  the Yukawa beta functions are, at one loop,
\bea
 \beta_{\lambda}=\beta^0_{\lambda}+\frac{6 y_t^2 \lambda}{16\pi^2} \quad , \quad \beta_{y_t}=\beta^0_{y_t}+\frac{6 y_t \lambda^2}{16\pi^2}
\label{eq:betafunctions}
\eea
where $\beta^0$ is a MSSM-like top Yukawa beta function. We immediately see that the parameter that controls the running of the Yukawas is $\abs{\lambda}$. Fixing this parameter, the ratio of the $A$-term over the soft mass is maximized when $\lambda_k$ and $\frac{X_k}{M_k}$ are parallel vectors in $k$ space. This leads to the bound
\begin{equation}
\abs{\frac{y_t A_t}{\delta m_{H_u}}} \leq \abs{\lambda}
\label{eq:littleAmh}
\end{equation}
where to retain perturbativity $\lambda$ needs to be of order one or smaller.  This bound is valid for the most general renormalizable superpotential that couples messengers with the Higgs at tree level. A similar bound relating the squark mass to the $A$-term can be obtained for the non-MFV model of section \ref{sec:NMFV models}. Note that the bound is independent of the messenger number. For messengers at $250$ TeV and $\lambda=1$ as considered in section \ref{sec:NMFV models} a Landau pole is obtained at $\sim 10^{10}$ GeV. A coupling $\lambda=3$ as considered in section \ref{sec:MFV} leads to a Landau pole less than a decade above the messenger scale.

\providecommand{\href}[2]{#2}\begingroup\raggedright\endgroup

\end{document}